\documentclass[epj,nopacs]{svjour}
\usepackage{array, eucal, mathtools, xfrac}
\usepackage{booktabs, array}
\usepackage{amsmath}%
\usepackage{amsfonts}%
\usepackage{amssymb}%
\usepackage{graphicx}
\usepackage{mathtools}
\begin{document}\sloppy
\title{A regularised singularity approach to phoretic problems}

\author{Thomas D. Montenegro-Johnson\inst{1} \and S\'{e}bastien Michelin\inst{2}
    \and Eric Lauga\inst{1}
}                     % Do not remove
\institute{Department of Applied Mathematics and Theoretical Physics, University
    of Cambridge, CMS, Wilberforce Rd, Cambridge, CB3 0WA, UK\and LadHyX -
D\'{e}partement de M\'{e}canique, Ecole polytechnique - CNRS, 91128 Palaiseau
Cedex, France}

\date{\today}

\abstract{An efficient, accurate, and flexible numerical method is proposed for the
solution of the swimming problem of one or more autophoretic particles in the
purely-diffusive limit. The method relies on successive boundary element
solutions of the Laplacian and the Stokes flow equations using regularised
Green's functions for swift, simple implementations, an extension of the
well-known method of ``regularised stokeslets'' for Stokes flow problems. The
boundary element method is particularly suitable for phoretic problems, since no
quantities in the domain bulk are required to compute the swimming velocity. For
time-dependent problems, the method requires no re-meshing and simple boundaries
such as a plane wall may be added at no increase to the size of the linear
system through the method of images. The method is validated against two
classical examples for which an analytical or semi-analytical solution is known,
a two-sphere system and a Janus particle, and provides a rigorous computational
pipeline to address further problems with complex geometry and multiple bodies.
%\PACS{
%      {PACS-key}{discribing text of that key}   \and
%      {PACS-key}{discribing text of that key}
%     } % end of PACS codes
} %end of abstract
\maketitle
\section{Introduction}
\subsection{Background}

Artificial self-propelled synthetic devices, or swimmers, at microscopic scales
have recently received increasing study, motivated both by advances in the
understanding of biological locomotion~\cite{lauga2009} and potential
engineering or biomedical applications~\cite{nelson2010microrobots}. Such
synthetic systems can be roughly divided into two main categories, depending on
whether they are externally actuated (e.g. by a rotating magnetic
field~\cite{ghosh2009,godinez2015complex}) or fuel-based~\cite{ebbens2010}. 

The principal advantage of fuel-based swimmers is that they utilise properties
of the immediate environment, rather than an externally-applied driving
mechanism, in order to generate propulsion. An important class of fuel-based
swimmers are those that exploit phoretic mechanisms, e.g. Janus
particles~\cite{walther2008janus}. The ability of phoretic swimmers to
self-propel depends on two chemical properties of their surface: \emph{mobility}
and \emph{activity}. Interaction with the solute species in the fluid within a
thin interaction layer leads to local pressure imbalances. A net slip flow (and
locomotion) develops at their boundary due to local solute concentration
gradients (diffusiophoresis), temperature gradients (thermophoresis) or electric
field (electrophoresis)~\cite{anderson1989}; a property of the surface
generically termed as \emph{mobility}. In order to create the required
gradients, the particle must also be \emph{active}, i.e.  modify the local field
gradients through chemical reaction catalysed at its surface or heat emission.
Canonical experimental phoretic systems include bi-metallic Au-Pt rods or
spheres~\cite{paxton2004} and PS-Pt coated spheres~\cite{howse2007}. Whilst the
precise details of the physico-chemical mechanisms underlying these systems are
still under debate~\cite{brown2014}, they have garnered much interest,
particularly as a means to study collective dynamics at the
micron-scale~\cite{theurkauff2012}. 

The recently-proposed continuum framework for self-diffusiophoretic
particles~\cite{golestanian2007designing} relies on three assumptions: (i) the
solute particle interaction layer is infinitely thin, and thus phoretic effects
are incorporated through a surface slip velocity; (ii) particle length scales
are microscopic, so that the P\'{e}clet number is zero and (iii) the catalytic
portion of the surface absorbs or releases chemical at a fixed rate.  Within
this framework, obtaining the particle swimming velocity requires solution of
the diffusive solute dynamics, and the boundary-driven Stokes flow problem in
the fluid domain. Hitherto, detailed  modelling of phoretic particles has
largely been analytical, focusing on simple
geometries~\cite{golestanian2007designing,shklyaev2014,michelin2015autophoretic}.
Numerical studies have typically used non-continuum methods for individual
particles~\cite{reigh2015catalytic}, or considered collective or non-trivial
confinement after much simplification of the solute-fluid
dynamics~\cite{sciortino2010numerical,ghosh2013self,soto2014}. {
    In contrast, boundary element techniques, which solve the solute-fluid
    dynamics directly, are only beginning to be
utilised~\cite{uspal2014self,singh2015many,michelin2015geometric}.}  

In this paper, we develop a versatile numerical framework based on boundary
integral methods and a regularised solution to the Laplace equation, inspired by
classical work on regularised solutions to Stokes flow
\cite{cortez2001method,Cortez05}. After a presentation of the approach and
numerical techniques, the method is validated against analytical solutions for a
two-sphere system and a classic Janus particle.

\subsection{Problem formulation}

Following previous
studies~\cite{michelin2015autophoretic,michelin2015geometric}, we consider the
purely diffusive limit of the self-diffusiophoretic locomotion problem: a rigid
particle of typical size $L$ achieves self-propulsion through its interaction
with a diffusing solute species present in the fluid. Neglecting solute
advection, the solute concentration completely decouples from the flow dynamics
and satisfies the Laplace equation,
\begin{equation}\label{eq:laplace}
D\nabla^2 c=0.
\end{equation}
The phoretic particle's chemical properties are characterized by the activity
$\mathcal{A}(\mathbf{x})$ and mobility $\mathcal{M}(\mathbf{x})$ of its surface
$S$, which may be homogeneous~\cite{michelin2015autophoretic} or
spatially-dependent, as for Janus particles~\cite{walther2008janus}. Hence, at
the particle's surface, solute is released or absorbed with flux,
\begin{equation}
    D\mathbf{n}\cdot\boldsymbol{\nabla}{c} = -\mathcal{A}(\mathbf{x}).
\end{equation}
Local gradients of solute concentration at the surface result in imbalances in
the pressure field that drive a flow within the interaction layer, where the
solute-particle interactions dominate. In the thin-interaction-layer limit, this
is equivalent to a net phoretic slip velocity~\cite{anderson1989}
\begin{equation}\label{eq:mobility}
    \mathbf{u} = \mathcal{M}(\mathbf{x})(\mathbf{I} - \mathbf{n}\mathbf{n})\cdot
    \boldsymbol{\nabla} c \quad \mbox{on } S,
\end{equation}
that in turn drives a fluid flow around the particle. Because of the micrometric
size of such particles, the flow dynamics is governed by Stokes flow equations
\begin{equation}
    \eta\nabla^2\mathbf{u} = \boldsymbol{\nabla}p, \quad
    \boldsymbol{\nabla}\cdot\mathbf{u} = 0,
    \label{eq:stokes_flow}
\end{equation}
for which fluid and solid inertia are negligible when compared with viscous
stresses. For freely-swimming particles, the total hydrodynamic force and torque
must vanish, which closes the system of equations for the translational and
angular velocities ($\mathbf{U}$, $\boldsymbol\Omega$) of the particle. In the
following, the problem is non-dimensionalised using $L$, $\mathcal{A}L/{D}$ and
$\mathcal{AM}/{D}$ as characteristic length, concentration and velocity scales,
respectively.

\section{Method}

A popular numerical means of solving the Laplace and Stokes flow equations is
the boundary element method, which effectively replaces the partial differential
equations \eqref{eq:laplace} and \eqref{eq:stokes_flow} over a three-dimensional
domain by integral equations over the particle's boundary. This reduction makes
the boundary element method particularly suitable for phoretic problems, which
are completely characterized by boundary values of the different fields and do
not require an explicit description of the bulk
distributions~\cite{michelin2015geometric}; indeed the coupling between the
solute and Stokes flow problems, eq.~\eqref{eq:mobility}, only requires the
value of the concentration along the boundary. Furthermore, since the
three-dimensional domain is not meshed, no re-meshing is required when analysing
the trajectories of several rigid particles.

\subsection{Finding the concentration: regularised sourcelets}
We propose a boundary element method to solve Laplace's equation inspired by the
regularised stokeslet method for Stokes flow. To this end, regularised free
space Green's functions are defined that satisfy
\begin{subequations}    
\label{eq:diff_gr_fun}
\begin{align}
    \nabla^2 G^\epsilon(\mathbf{x} - \mathbf{x}_0) &= -\phi_\epsilon(\mathbf{x} -
    \mathbf{x}_0), \\ 
    \nabla^2 \mathbf{K}^\epsilon(\mathbf{x} - \mathbf{x}_0) &= -\boldsymbol{\nabla}
    \phi_\epsilon(\mathbf{x} - \mathbf{x}_0),
    \end{align}
\end{subequations}
where $\phi_\epsilon$ is a `blob' source located at $\mathbf{x}_0$, with small
regularisation parameter $\epsilon$. { In the standard boundary
element method, the use of singular Green's functions necessitates careful
treatment of surface integrals. This regularisation allows the use of simple,
standard quadrature routines for all surface integrals in the regularised
boundary integral equation~\cite{Smith09b}.} As is common for the method of
regularised stokeslets~\cite{Cortez05}, we choose a regularisation function of
the form 
\begin{equation}
    \phi_\epsilon(\mathbf{x} -
    \mathbf{x}_0) = \frac{15\epsilon^4}{8\pi r_\epsilon^7}, \quad r_\epsilon^2 =
    r^2 + \epsilon^2,
    \label{eq:blob_choice}
\end{equation}
where $r = |\mathbf{x} - \mathbf{x}_0|$. For this choice of regularisation
function, the solutions to eqs~\eqref{eq:diff_gr_fun} are
\begin{subequations}    \label{eq:reg_gf}
\begin{align}
    G^\epsilon(\mathbf{x} - \mathbf{x}_0) &= -\frac{2r^2 + 3\epsilon^2}{8\pi
    r_\epsilon^3}, \\
    K^\epsilon_j(\mathbf{x} - \mathbf{x}_0) &= r_j\frac{2r^2 +
    5\epsilon^2}{8\pi r_\epsilon^5},
\end{align}
\end{subequations}
with $r_j = (\mathbf{x} - \mathbf{x}_0)_j$. Note that for $r \neq 0$, as
$\epsilon \to 0$ we recover the well-known fundamental solutions of a point sink and source dipole, respectively,
\begin{subequations}
\begin{alignat}{2}
    \lim_{\epsilon\to 0} G^\epsilon(\mathbf{x},\mathbf{x}_0) &= 
    G^0(\mathbf{x},\mathbf{x}_0) &&= -\frac{1}{4\pi r}, \\
    \lim_{\epsilon\to 0} K^\epsilon_j(\mathbf{x},\mathbf{x}_0) &=
    K^0_j(\mathbf{x},\mathbf{x}_0) &&= \frac{r_j}{4\pi r^3}\cdot
\end{alignat}
\end{subequations}
Following the method of regularised stokeslets, these regularised Green's functions are used to construct
a regularised boundary integral equation for the diffusion equation,
\begin{align}
    \int_{V} c(\mathbf{x}) \phi_\epsilon(\mathbf{x} - \mathbf{x}_0)\,\mathrm{d}V_{x} =
    \int_{S}&\Big[c(\mathbf{x})\mathbf{K}^\epsilon(\mathbf{x},\mathbf{x}_0) 
\cdot\mathbf{n}(\mathbf{x}) \nonumber \\
&- \frac{\partial
    c(\mathbf{x})}{\partial\mathbf{n}}G^\epsilon (\mathbf{x},\mathbf{x}_0) 
    \Big ]\,\mathrm{d}S_{x},
    \label{eq:diff_bem}
\end{align}
where $\mathbf{n}$ is the normal to the swimmer surface pointing into the fluid,
and $\mathbf{x}_0$ is a fixed point. In the singular
boundary integral method, the left-hand side term is computed exactly
\begin{equation}
    \int_V c(\mathbf{x}) \delta(\mathbf{x} - \mathbf{x}_0)\,\mathrm{d}V_{x}
    = \lambda c(\mathbf{x}_0),
\end{equation}
where $\lambda = 0,1/2,1$ depending on whether the evaluation point
$\mathbf{x}_0$ is inside, on, or outside the boundary, respectively. In the
regularised method, this critical step is less straightforward. Firstly, assuming
slow variations of the concentration field at the scale of $\epsilon$, we obtain
\begin{equation}
    \int c(\mathbf{x}) \phi_\epsilon(\mathbf{x} - \mathbf{x}_0)\,\mathrm{d}V_{x}
    \approx c(\mathbf{x}_0)\int\phi_\epsilon(\mathbf{x} -
    \mathbf{x}_0)\,\mathrm{d}V_{x},
\end{equation}
which is similar to the singular case for flat surfaces. Then, the integral on
the right-hand side must be evaluated. When the mean curvature of the surface is
negligible in comparison to $\epsilon^{-1}$, the approximation $\lambda = 0$
(inside), $1/2$ (on the surface) and $1$ (outside) provides sufficient accuracy.
Note that here, inside/outside (resp. on the surface) is understood as much
further (resp. closer) from the surface than $\epsilon$. For a smooth surface,
the accuracy can be improved by taking into account the curvature of the
particle's surface through a Taylor's series expansion of the surface's
geometry. At leading order, the value of $\lambda$ for an evaluation point on
the boundary becomes
\begin{equation}
\lambda\approx\frac{1}{2}+\frac{\kappa \epsilon}{4},
\end{equation}
where $\kappa$ is the mean local curvature of the surface, counted positively
when the particle is locally convex. For a sphere with $\kappa\epsilon = 0.01$,
the integral of $\phi_\epsilon(\mathbf{x} - \mathbf{x}_0)$ can be computed
exactly, and the above expansion is accurate to $\mathcal{O}(10^{-8})\%$.
However, this approximation clearly breaks down near a cusp or a corner, where
additional care should be taken.

It is a relatively simple matter to adjust the boundary integral formulation to
include simple confining boundaries. For instance, the presence of a plane wall
or free-surface may be modelled by replacing the Green's
functions~\eqref{eq:reg_gf} in eq.~\eqref{eq:diff_bem} with the appropriate
half-plane solutions.  Crucially, this substitution results in no increase to
the size of the resultant linear system, and multi-particle systems over a
boundary may be modelled as readily as in free space. Such images are commonly
employed for point singularities in electrostatics, and also for studying
viscous swimmers and cilium-induced flow~\cite{gueron1992}. However, in
contrast with regularised image systems for Stokes
flows~\cite{ainley2008method}, the half-space Green's functions for the
Laplacian above a plane boundary located at $z = 0$ are remarkably simple, 
\begin{subequations}
\begin{alignat}{2}
    G^{\epsilon_w}(\mathbf{x} - \mathbf{x}_0) &= G^{\epsilon}(\mathbf{x} -
    \mathbf{x}_0) &&+ \beta G^{\epsilon}(\mathbf{x} - \mathbf{x}_{\mathrm{im}}), \\
    K^{\epsilon_w}_j(\mathbf{x} - \mathbf{x}_0) &=  K^{\epsilon}_j(\mathbf{x} -
    \mathbf{x}_0) &&+ \beta \gamma_j K^{\epsilon}_j(\mathbf{x} -
    \mathbf{x}_{\mathrm{im}}),
    \label{eq:reg_gf_zwall}
\end{alignat}
\end{subequations}
where $\mathbf{x}_{\mathrm{im}} = (x_0,y_0,-z_0)$ and $\gamma_{1} = \gamma_2 =
1, \gamma_3 = -1$. The constant $\beta$ takes the value $\beta = -1$ for
absorbing boundaries, $c(x,y,0) = 0$, and $\beta = +1$ for no flux boundaries
$\partial c/\partial z = 0$.

\subsection{Calculating the slip velocity}

Once the surface concentration has been obtained, its gradient
along the surface can be calculated through evaluation
of the integral,
\begin{align}
    \lambda \boldsymbol{\nabla} c(\mathbf{x}_0) =
    \int_{S}&c(\mathbf{x})\mathbf{L}^\epsilon(\mathbf{x},\mathbf{x}_0) 
\cdot\mathbf{n}(\mathbf{x}) \nonumber \\
&- \frac{\partial
    c(\mathbf{x})}{\partial\mathbf{n}}\mathbf{K}^\epsilon (\mathbf{x},\mathbf{x}_0) 
    \,\mathrm{d}S_{x},
    \label{eq:slip_eqn}
\end{align}
which can be formally obtained by differentiating eq.~\eqref{eq:diff_bem} with
respect to $\mathbf{x}_0$. {The concentration gradient
$\boldsymbol{\nabla} c(\mathbf{x})$ is then substituted into
eq.~\eqref{eq:mobility} to give the surface slip velocity.} Here, we
have neglected an additional left hand side term $c(\mathbf{x}_0) \nabla
\lambda(\mathbf{x}_0)$: we are only interested in the part of these gradients
tangential to the surface, and along smooth surfaces $\lambda$ changes very
slowly. In free space, the Green's function
$\mathbf{L}^\epsilon(\mathbf{x},\mathbf{x}_0)$ is the derivative of the dipole
\begin{align}
    \frac{\partial K_i^\epsilon}{\partial x_j} = L^\epsilon_{ij} 
    &= \delta_{ij} \frac{2r^2 + 5\epsilon^2}{8\pi r_\epsilon^5}
    \nonumber \\
    &+ r_i r_j\left( \frac{1}{2\pi r_\epsilon^5} - \frac{10r^2 +
    25\epsilon^2}{8\pi r_\epsilon^7} \right),
\end{align}
where $\delta_{ij}$ is the Kronecker delta tensor. Once eq.~\eqref{eq:diff_bem}
has been solved to obtain the concentration field, $c(\mathbf{x})$ and
${\partial c(\mathbf{x})}/{\partial\mathbf{n}}$ are known on the particle's
boundary. Therefore, computing the {concentration
gradient~\eqref{eq:slip_eqn}} does not require solving another integral
equation, or inverting a linear system. The singularity is an order higher than
in the concentration calculation, { however the regularisation
ensures that the Green's functions remain integrable; either higher-order
quadrature or increased regularisation is sufficient for convergence. If the value
of the regularisation parameter $\epsilon$ is increased, $\lambda$ too must be
recalculated for the gradient evaluation only.} We have employed this increased
regularisation in the results that follow. For flat surfaces, it is worth noting
that an alternative simple approach to calculating the slip velocity can be
obtained through finite differences of interpolated values of the surface
concentration. 

\subsection{Finding the surface traction: regularised stokeslets}

To solve for the particle's swimming velocity and the induced fluid flow, we
employ the regularised stokeslet boundary element
method~\cite{Cortez05,Smith09b}. This method has the advantage that finding the
velocity of the particle only requires the tractions on the particle surface,
and not the flow field in the bulk. Furthermore, much of the numerical code
employed to solve the Laplace problem can be reused, with a substitution of the
appropriate Green's functions. 

For Stokes flow, eqs~\eqref{eq:stokes_flow}, the fluid velocity at any point in the
domain can be evaluated through integrals of stokeslets $\mathbf{S}$ and
stresslets $\mathbf{T}$ over the surface of the particles,
\begin{align}
    \lambda u_j(\mathbf{x}_0) = \int_S
    &S_{ij}^\epsilon(\mathbf{x},\mathbf{x}_0)f_i(\mathbf{x})
    \nonumber \\
    &-u_i(\mathbf{x})T_{ijk}^\epsilon(\mathbf{x},\mathbf{x}_0)
    n_k(\mathbf{x})\,\mathrm{d}S_x,
    \label{eq:reg_bem}
\end{align}
where $\lambda, \mathbf{x}_0$ and $\mathbf{n}$ are as in
eq.~\eqref{eq:diff_bem}. We utilise the same regularisation as for the
Laplacian~\eqref{eq:blob_choice}, for which we have regularised stokeslet
$S_{ij}^\epsilon$ and stresslet $T_{ijk}^\epsilon$~\cite{Cortez05}
\begin{subequations}
\begin{align}
    S_{ij}^\epsilon(\mathbf{x},\mathbf{x}_0) &= \frac{\delta_{ij}(r^2 +
    2\epsilon^2) + r_i r_j}{r_\epsilon^3}, \\
    T_{ijk}^\epsilon(\mathbf{x},\mathbf{x}_0) &= -\frac{6r_i r_j
    r_k}{r_\epsilon^5} \nonumber \\
    &\phantom{={ }}-\frac{3\epsilon^2\left(r_i\delta_{jk} + r_j\delta_{ik} + 
    r_k\delta_{ij}\right)}{r_\epsilon^5}\cdot
\end{align}
\end{subequations}
{Substituting $\mathbf{u}(\mathbf{x}) = \mathbf{u}^s(\mathbf{x}) +
    \mathbf{U} + \boldsymbol{\Omega}\wedge(\mathbf{x} - \mathbf{x_c})$ for
    surface slip velocity $\mathbf{u}^s$, swimming translational and angular
    velocities $\mathbf{U},\boldsymbol{\Omega}$, and particle centre
    $\mathbf{x_c}$ into eq.~\eqref{eq:reg_bem}, we see
\begin{align}
    &\lambda u^s_j(\mathbf{x}_0) + \int_S u^s_i(\mathbf{x})T_{ijk}^\epsilon(\mathbf{x},\mathbf{x}_0)
    n_k(\mathbf{x})\,\mathrm{d}S_x \nonumber \\
    &= \int_S S_{ij}^\epsilon(\mathbf{x},\mathbf{x}_0)f_i(\mathbf{x})
    \nonumber \\
    &-[U_i + (\boldsymbol{\Omega}\wedge[\mathbf{x}_0 - \mathbf{x}_c])_i]T_{ijk}^\epsilon(\mathbf{x},\mathbf{x}_0)
    n_k(\mathbf{x})\,\mathrm{d}S_x,
    \label{eq:swim_bem}
\end{align}
where the surface tractions $\mathbf{f}$ and swimming velocities
$\mathbf{U},\boldsymbol{\Omega}$ are unknowns. The linear system is closed for the
unknown swimming velocities by enforcing the zero net force and torque
conditions
\begin{equation}
    \int_S \mathbf{f}(\mathbf{x})\,\mathrm{d}S_x = \mathbf{0},
    \quad
    \int_S (\mathbf{x} - \mathbf{x}_c)\wedge\mathbf{f}(\mathbf{x})\,\mathrm{d}S_x = 
    \mathbf{0},
\end{equation}
as constraints in the matrix system~\cite{ishikawa2006hydrodynamic} that arises from the discretisation of
eq.~\eqref{eq:swim_bem}.}

\subsection{Numerical implementation}

We now discuss the numerical implementation of this framework, which is freely
available {as the package ``RegBEM Phoretic''} from the Matlab file
exchange~\cite{monjohn}.

\begin{figure}[tb]
\begin{center}
\includegraphics[scale=1]{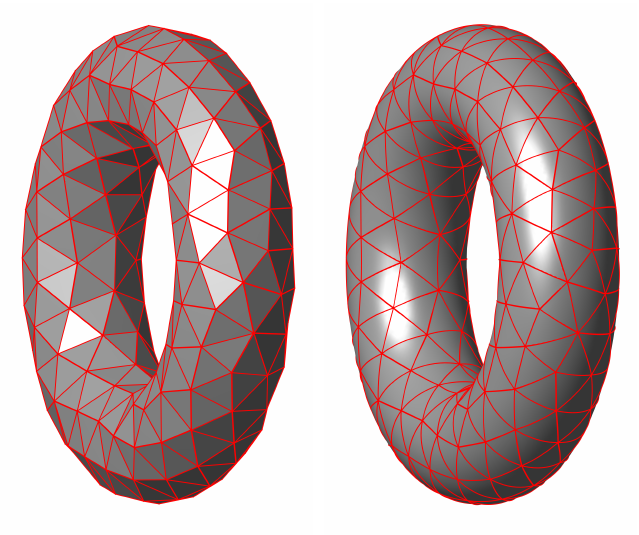}
\end{center}
\caption{The surface mesh description. The geometrically quadratic mesh (top
    right) provides a much more faithful representation of the particle surface
    than a linear mesh with the same number of elements (top left).}
\label{fig:linear_vs_quad_geom}
\end{figure}

The first step in solving the phoretic locomotion problem with the boundary
element method is to define the computational mesh of the particle surfaces.
This part of the procedure is essentially distinct from the boundary element
solver, and many automatic mesh generation packages are available. We use the
Matlab package DistMesh~\cite{persson2004simple} to generate a flat triangular
representation of our surfaces, and then custom routines are used to extend this
to a piecewise quadratic surface description
(fig.~\ref{fig:linear_vs_quad_geom}). A quadratic geometric representation of
the surface allows the geometry to be better captured by fewer elements, and
gives a significant increase in solution accuracy~\cite{frijns2000improving}.
Integration of the Green's functions over each element is performed numerically
using adaptive Fekete quadrature~\cite{taylor2000algorithm,smith2012symmetry},
with lookup table routines written by John Burkardt~\cite{burkardt}. The
coordinates of these quadrature points are projected from the canonical triangle
to each quadratic element using the routines \texttt{abc} and
\texttt{interp\char`_p} from the BEMLIB library~\cite{pozrikidis2002practical},
adapted for use in Matlab.

Whilst the mesh is geometrically quadratic, the discretisation of the
concentration $c(\mathbf{x})$ and the surface tractions $\mathbf{f}(\mathbf{x})$
can be defined independently, according to the user's needs. A common
implementation of the boundary element method is with constant panels; the
unknown field on the surface is discretised as constant over each element,
collocated at the element centroid \cite{Smith09b,smith2012symmetry}. The
constant panel method has a simple implementation, and can provide a relatively
high level of accuracy. However, phoretic problems involve solving a series of
subproblems: finding an unknown surface concentration, evaluating the concentration
gradient to give a slip velocity, and then finding the unknown surface tractions
and propulsion velocity arising from this surface slip. Achieving high accuracy
at each step is essential in order to avoid propagating errors. Thus we
employ a linear panel boundary element method, where the surface unknowns are
discretised as linear over each element. 

The linear panel method confers a number of advantages over the constant panel
method. Firstly, it provides a higher-order, more accurate representation of the
unknown surface field.  Secondly, each element has exactly three linear nodes
where the unknowns are collocated, the vertices of the triangle, but for closed
surfaces each vertex is shared between $6$ elements, on average.  Thus, for any
given mesh the number of degrees of freedom for the linear panel method is
approximately half that of the constant panel mesh, reducing the time to solve
the linear system by approximately a factor of $8$.  Finally, Fekete quadrature
points are clustered near the element vertices, leading to more accurate
evaluation of the nearly singular element integrals for any given number of
quadrature points, when compared to the constant panel method.

The element-based discretisation of the concentration and surface tractions
ensures that the solution space is decoupled from the numerical quadrature used
to evaluate the surface integrals. Whilst more complicated than traditional
implementations of regularised stokeslets~\cite{Cortez05,guo2014cilia}, this
boundary element implementation has the advantage of maintaining accuracy whilst
greatly reducing the size of the linear system to be solved~\cite{Smith09b}.
Furthermore, it allows for adaptive quadrature to be used for integrals over
elements far from the collocation point, significantly reducing the time
required to assemble the matrix system. The simulations in the following section
typically run in around a minute on a 16GB Macbook Retina Pro (2013), though if
less accuracy is required simulation time may be of the order of seconds.

\section{Validation of numerical scheme}

We now proceed with a validation of our
method against well-established analytical results for a two-sphere system
\cite{michelin2015autophoretic}, and a Janus particle
\cite{golestanian2007designing}. The two sphere system has non-trivial geometry
with simple surface chemistry resulting in constant flux. In contrast, the Janus
particle has a simple geometry with discontinuous surface chemistry, resulting
in zero-flux and unit flux hemispheres.

\subsection{A two-sphere system}
\begin{figure}[tb]
    \begin{center}
	\includegraphics[scale=1]{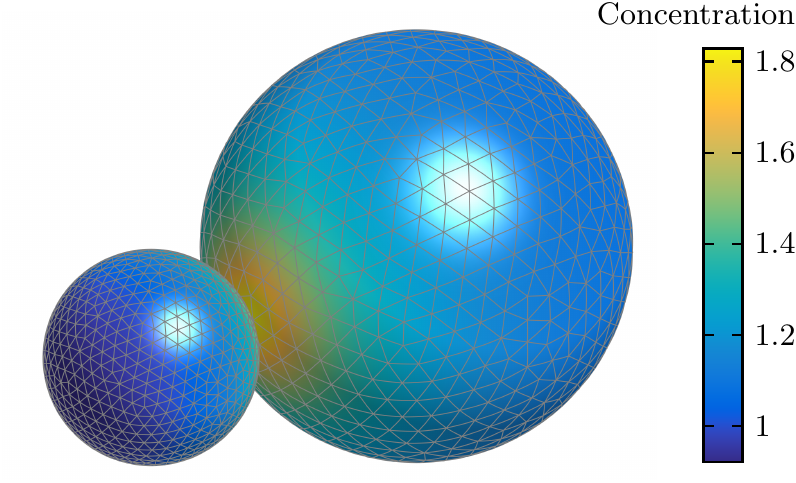}    
    \end{center}
    \caption{Surface concentration of the two-sphere system with $R_2 = 0.5$ and $d_c = 0.1$.}
    \label{fig:two_sphere_conc}
\end{figure}

We first validate our method against the semi-analytical solution of the
phoretic problem considered in Ref.~\cite{michelin2015autophoretic}; a swimmer comprising two
chemically-homogeneous spheres with radii ($R_1=1,R_2$), rigidly linked and
separated by a contact distance $d_c$ (Figure~\ref{fig:two_sphere_conc}). The
performance of the code is examined by comparing the surface concentration, slip
velocity and swimming velocity with the analytical result. The following
computations were performed with $1320$ linear nodes per sphere, with
regularisation $\epsilon = 0.002R_{1,2}$ for the concentration and traction
calculations, and $\epsilon = 0.01R_{1,2}$ for the slip velocity evaluation. 

\begin{figure}[tb]
    \centering
    \includegraphics[scale=1]{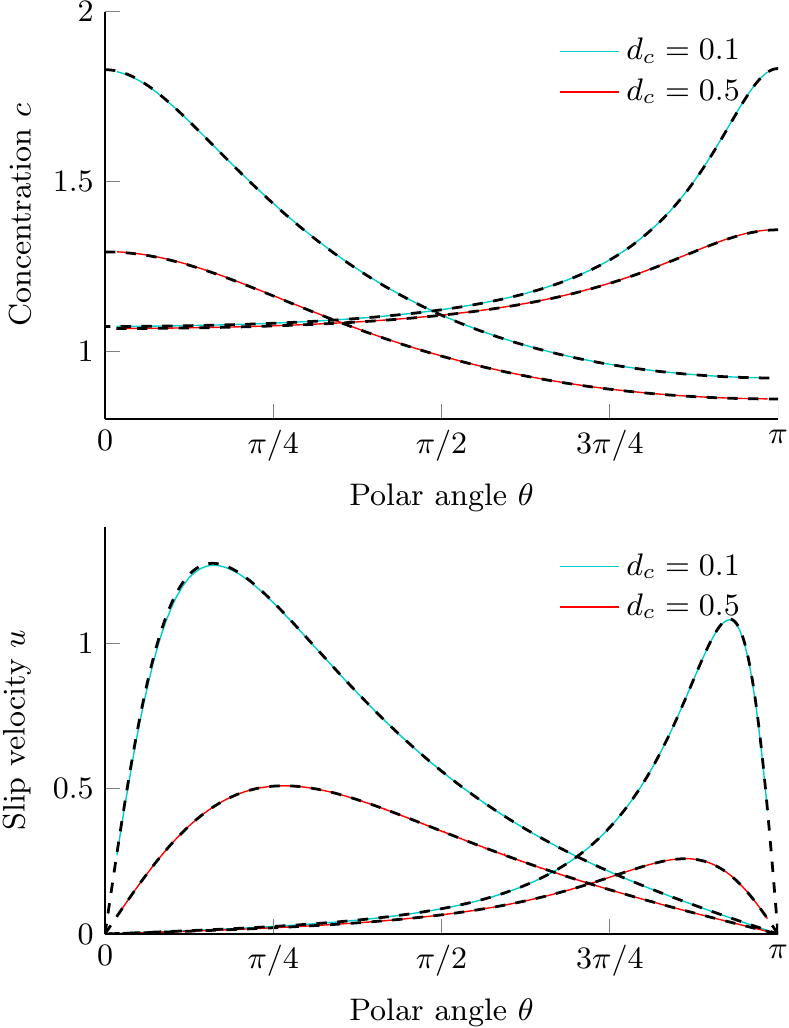}
    \caption{Surface concentration (upper) and slip velocity (lower) of the
two-sphere system, for $d_c = 0.1$ (cyan) and $d_c = 0.5$ (red). Sphere $1$
takes its peak values on the right of the plot, and the smaller sphere $2$ on
the left. The analytical result~\cite{michelin2015autophoretic} is virtually
indistinguishable, and shown for reference in black.}
    \label{fig:two_sphere_conc_and_slip}
\end{figure}

The surface concentration and slip velocity as a function of the polar angle
$\theta$ from the midline between the spheres is shown in
figure~\eqref{fig:two_sphere_conc_and_slip}. The solution is given for
almost-touching spheres $d_c = 0.1$ and for $d_c = R_2 = 0.5$, and compared to
the analytical result. The mean relative errors for $d_c = 0.1$ on the second
sphere are $0.05\%$ for the surface concentration and $0.2\%$ for the slip
velocity, while for $d_c = 0.5$ the mean relative errors are $0.05\%$ and
$0.2\%$, showing excellent agreement with analytical values.

\begin{figure}[tb]
    \includegraphics[scale=1]{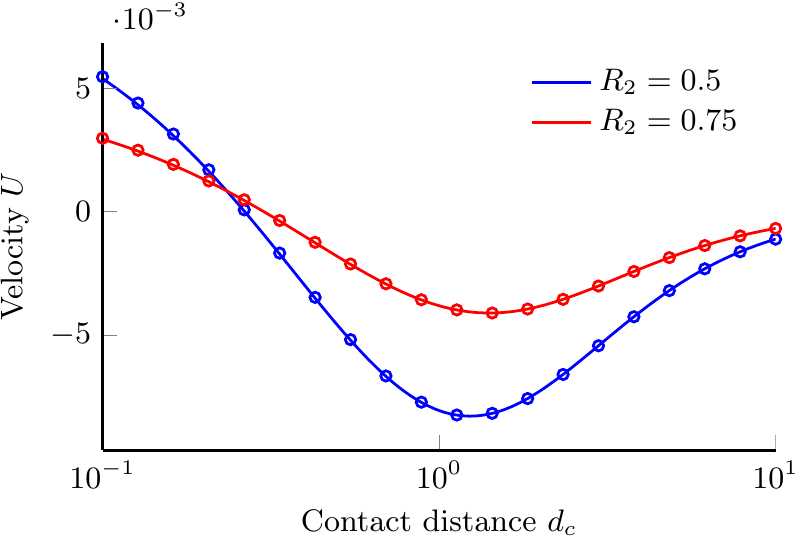}
    \caption{Self-propulsion velocity of a two-sphere rigid system as function of the
        contact distance $d_c$, with $R_1 = 1$, $R_2 = 0.5$ (blue) and $R_2 =
        0.75$ (red), obtained numerically using the present approach (crosses) and analytically~\cite{michelin2015autophoretic} (solid lines).}
    \label{fig:two_sphere_vel}
\end{figure}

The self-propulsion velocity of the two-sphere assembly as a function of contact
distance $d_c$ is then computed, for $R_2 = 0.5$ and $R_2=0.75$, and compared to
the results of Michelin and Lauga~\cite{michelin2015autophoretic}; the numerical
results are in good agreement with the analytical prediction
(Figure~\ref{fig:two_sphere_vel}). Because the swimming velocity may vanish for
certain distances, the relative error is evaluated here by dividing the
absolute error by the average of the absolute speed in our sample, obtaining a
maximum relative error of $1.5\%$ at the closest contact distance of $d_c =
0.1$, the system shown in figure~\ref{fig:two_sphere_conc}. For distances of
$\mathcal{O}(1)$ or greater, the relative error is less than $0.1\%$.

\subsection{Validation: A spherical Janus particle}

\begin{figure}[tb]
    \begin{center}
	\includegraphics[scale=1]{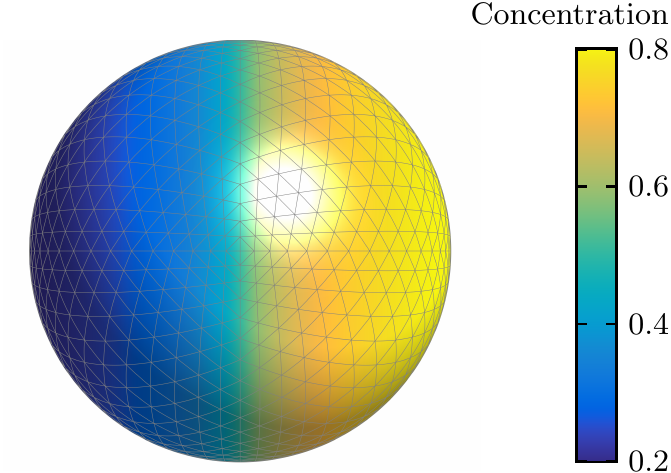}
    \end{center}
    \caption{Surface concentration of a Janus particle, calculated using our
    numerical method.}
    \label{fig:janus_conc}
\end{figure}

A more common route to breaking symmetry and achieving self-propulsion in
autophoretic system is chemical patterning: such Janus
particles~\cite{walther2008janus} have different chemical properties (i.e.
activity and mobility) on their two halves, and in particular may catalyze a
chemical reaction only on one of their hemispheres (fig.~\ref{fig:janus_conc}).
In such cases, the normal flux of solute into the fluid domain is discontinuous
at the edge of the coating, and it is important that our method is able to
handle such discontinuities.

For an axisymmetric, spherical Janus particle with uniform mobility
$\mathcal{M}=1$ and a hemispherical reactive cap with activity $\mathcal{A}=1$ (the other hemisphere being inert, $\mathcal{A}=0$),
the surface concentration is given by~\cite{golestanian2007designing,michelin2014phoretic}
\begin{equation}
    c(r,\mu) = \sum_{p=0}^\infty c_p(r)L_p(\mu), \quad c_p(r) =
    \frac{k_p}{(p+1)r^{p+1}},\label{eq:janus_an}
\end{equation}
where $L_p(\mu)$ is the $p$\textsuperscript{th} Legendre polynomial with $\mu =
\cos\theta$ the cosine of the polar angle, and the coefficients $k_p$ are obtained as
\begin{gather}
    k_0 = \frac{1}{2},\quad k_{2q} = 0 \nonumber\\
    k_{2q-1} = (-1)^{q+1}\frac{4q - 1}{4q - 2}\frac{(2q)!}{(2^qq!)^2}, \quad
    q\geq 1.
\end{gather}
The slip velocity can be easily obtained by taking the gradient of eq.~\eqref{eq:janus_an}. For a homogenous mobility, the swimming velocity is obtained in terms of $c$ as~\cite{michelin2013spontaneous}
\begin{equation}
U=-\int_{-1}^1\mu c(1,\mu)\mathrm{d} \mu=\frac{k_1}{3}=\frac{1}{4}\cdot
\end{equation}
The series solution for the concentration converges sufficiently quickly for a
quantitative comparison, however even with $200$ non-zero coefficients the
analytical slip velocity is only suitable for a qualitative comparison.

\begin{figure}[tb]
    \includegraphics[scale=1]{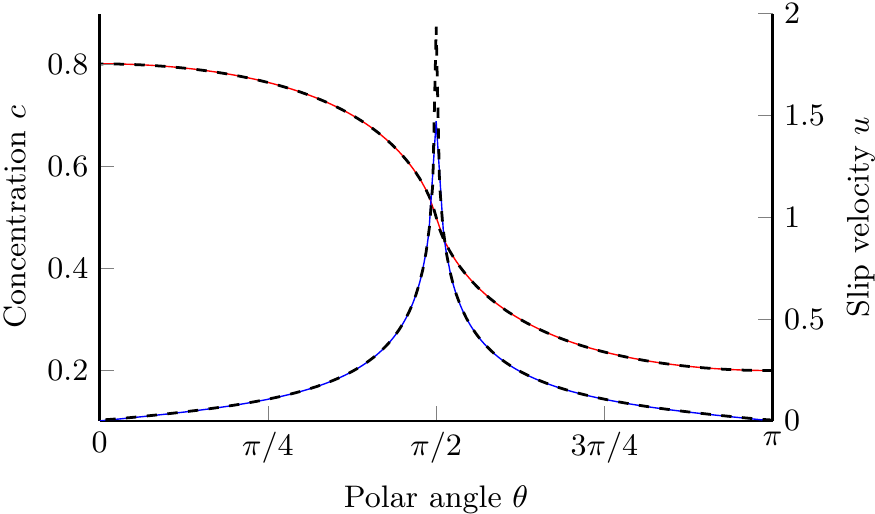}
    \caption{Surface concentration (red) and slip velocity (blue) of a Janus
        particle with uniform mobility and a hemispherical cap with unit
        activity. Analytical values calculated from the series are overlaid with
    dashed black lines.}
    \label{fig:janus_comb}
\end{figure}

The code was validated for regular meshes with $1026$ and $4098$ linear nodes,
for which the relative error in the calculated swimming velocity was $4.5\%$ and
$0.4\%$ respectively. The concentration field and slip velocity are shown
together in figure~\ref{fig:janus_comb} for the refined mesh. The mean relative
error in the surface concentration is $0.8\%$. While the cusp in the slip
velocity may not be fully captured by our model, this clearly does affect the
calculation of swimming velocities. Away from the cusp, the mean error in the
slip velocity is $0.7\%$, though due to the slow convergence of the series
a more accurate estimate is not possible.

\section{Discussion}

\begin{figure}[tb]
    \begin{center}
    \includegraphics[scale=1]{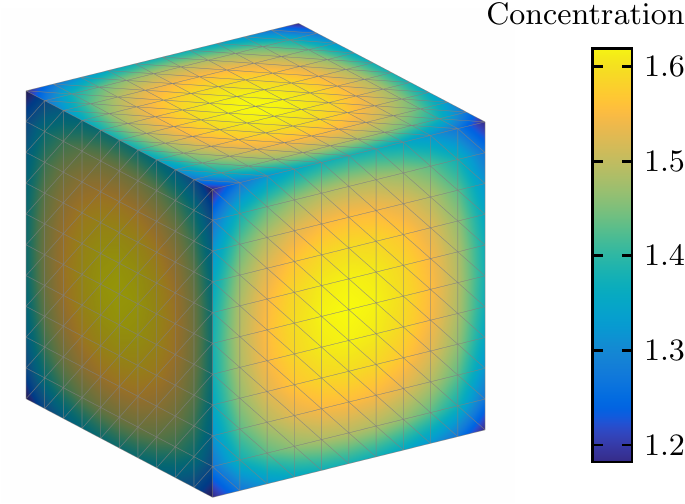}
\end{center}
    \caption{Surface concentration of a cube with uniform activity
$\mathcal{A}=1$ and sides of length $L = 2$. The double layer constant
$\lambda=1/2,3/7,7/8$ for the faces, edges, and corners respectively. }
    \label{fig:cube_conc}
\end{figure}

\begin{figure*}[tb]
    \begin{center}
    \includegraphics[scale=1]{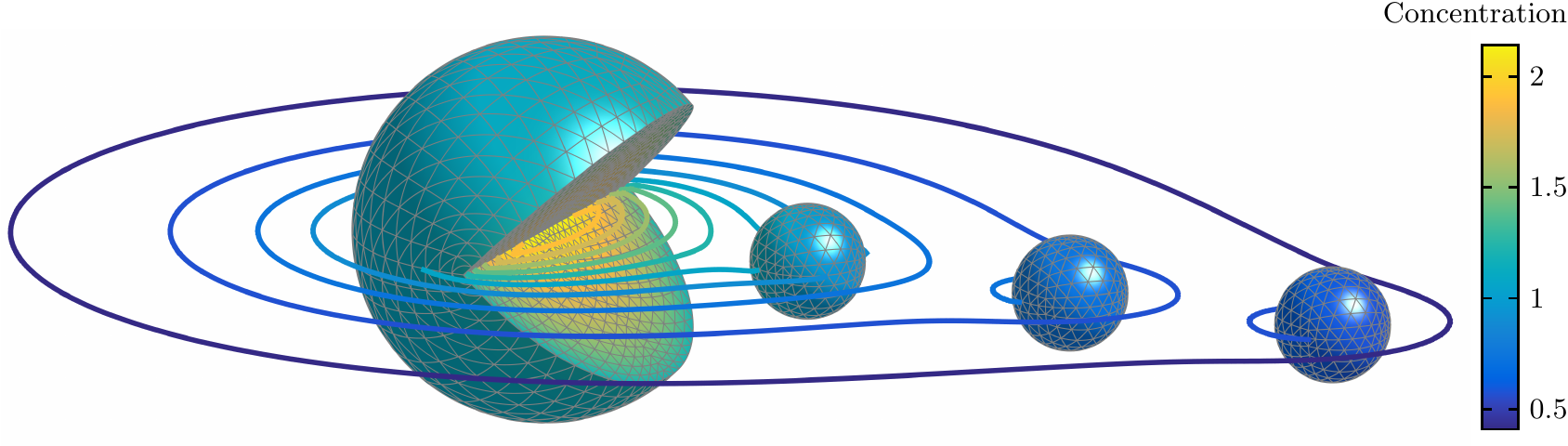}
\end{center}
    \caption{Surface concentration and contours of the concentration field in
the bulk for a non-trivial geometry with multiple particles: a phoretic pacman
interacting with phoretic pacdots. Activity is uniform, $\mathcal{A}=1$ for all
surfaces.}
    \label{fig:pacman}
\end{figure*}

In this work, we have proposed a generic and versatile pipeline to solve
phoretic motion problems in the classical continuum framework. The method relies
on successive boundary element solutions of the Laplacian, using regularised
sourcelets, and the Stokes flow equations using regularised
stokelets~\cite{Cortez05}. The method fully exploits the advantages of boundary
element methods for phoretic problems; the solute diffusion and Stokes flow
problems are only coupled on the boundary. The method has been validated against
analytical solutions for two separate phoretic systems, entailing multiple
particles and discontinuous flux conditions. In contrast with simulation
techniques requiring a computational mesh of the entire fluid domain, dynamic
simulations of multiple moving particles require no re-meshing between
time-steps, providing a computationally efficient approach. This method may
prove particularly useful to address problems with complex geometries that do
not readily admit an analytical solution; figure~\ref{fig:cube_conc} shows the
surface concentration of a cube with uniform activity. The approach could
equally be used to examine L-shaped particles \cite{ten2014gravitaxis} without
the slender body approximation. Furthermore, the method can be applied for
multiple particle systems with non-trivial geometries; figure~\ref{fig:pacman}
shows the surface concentration and midplane concentration contours of a
phoretic pacman of uniform activity, ready to exploit the concentration gradient
generated by his mouth to eat the uniformly active pacdots. 

\section{Acknowledgements}
TDMJ is supported by a Royal Commission for the Exhibition of 1851 Research
Fellowship. SM acknowledges the support of the French Ministry of Defense (DGA).
This work was funded in part by a European Union Marie Curie CIG Grant to EL.

\bibliographystyle{unsrt}
\bibliography{refs}

\end{document}